\pdfoutput=1
\documentclass[journal]{IEEEtran}
\usepackage[T1]{fontenc}
\usepackage{textcomp}
\usepackage{lmodern}
\usepackage[latin1]{inputenc}
\usepackage[swedish,english]{babel}
\usepackage{graphics}
\usepackage{amssymb}
\usepackage{amsmath}
\usepackage{graphicx} 
\usepackage{color} 
\usepackage{transparent} 
\usepackage{bm}
\usepackage{fmtcount}
\usepackage{multirow}
\usepackage{caption}
\usepackage{url}
\usepackage{epstopdf}
\usepackage{babel}
\makeatletter
\adddialect\l@ENGLISH\l@english
\makeatother

\DeclareGraphicsExtensions{.eps}

\ifCLASSINFOpdf
\else
\fi

\newcommand{\MYfooter}{\smash{\scriptsize
\hfil\parbox[t][\height][t]{\textwidth}{\centering
~\\
~}\hfil\hbox{}}}

\newcommand{\MYarxivheader}{\smash{\scriptsize
\hfil\parbox[t][\height][t]{\textwidth}{\centering
\copyright\ 2015 IEEE. Personal use of this material is permitted. Permission from IEEE must be obtained for all other uses, in any current or future media, including reprinting/republishing this material for advertising or promotional purposes, creating new collective works, for resale or redistribution to servers or lists, or reuse of any copyrighted component of this work in other works.}\hfil\hbox{}}}

\makeatletter

\def\ps@headings{%
\def\@oddhead{\mbox{}\scriptsize\rightmark \hfil \thepage}
\def\@evenhead{\scriptsize\thepage \hfil \leftmark\mbox{}}
\def\@oddfoot{\MYfooter}%
\def\@evenfoot{\MYfooter}}

\def\ps@IEEEtitlepagestyle{%
\def\@oddhead{\MYarxivheader}%
\def\@evenhead{\scriptsize\thepage \hfil \leftmark\mbox{}}%
\def\@oddfoot{\MYfooter}%
\def\@evenfoot{\MYfooter}}

\makeatother

\begin{document}

\title{Comprehensive and Macrospin-Based Magnetic Tunnel Junction Spin Torque Oscillator Model -- Part II: Verilog-A Model Implementation}

\author{Tingsu~Chen,~\IEEEmembership{Student Member,~IEEE,}
        Anders~Eklund,~\IEEEmembership{Student Member,~IEEE,}
        Ezio~Iacocca,~\IEEEmembership{Student Member,~IEEE,}
        Saul~Rodriguez,~\IEEEmembership{Member,~IEEE,}
        Gunnar~Malm,~\IEEEmembership{Senior Member,~IEEE,}
        Johan~$\AA$kerman,~\IEEEmembership{Member,~IEEE,}
        and Ana~Rusu,~\IEEEmembership{Member,~IEEE,} 
\thanks{Manuscript received October 20, 2014; revised December 5, 2014; accepted December 15, 2014. This research is supported by Swedish Research Council (VR).} 
\thanks{Tingsu Chen, Anders Eklund, Saul Rodriguez, Gunnar Malm and Ana Rusu are with the Department
of Integrated Devices and Circuits, KTH Royal Institute of Technology, 164 40 Kista, Sweden.
(e-mail: tingsu@kth.se, ajeklund@kth.se, saul@kth.se, gunta@kth.se and arusu@kth.se).} 
\thanks{Ezio Iacocca and Johan $\AA$kerman are with the Department of Physics, University of Gothenburg, 412 96 Gothenburg, Sweden.(e-mail: ezio.iacocca@physics.gu.se).}
\thanks{Johan $\AA$kerman is also with the Department of Materials and Nano Physics, KTH Royal Institute of Technology, 164 40 Kista, Sweden. (e-mail: akerman1@kth.se).}}

\markboth{IEEE TRANSACTIONS ON ELECTRON DEVICES}{T. Chen \MakeLowercase{\textit{et al.}}: Comprehensive MTJ STO model}

\maketitle

\begin{abstract}
The rapid development of the magnetic tunnel junction (MTJ) spin torque oscillator (STO) technology demands an analytical model to enable building MTJ STO-based circuits and systems so as to evaluate and utilize MTJ STOs in various applications. In Part I of this paper, an analytical model based on the macrospin approximation, has been introduced and verified by comparing it with the measurements of three different MTJ STOs. In Part II, the full Verilog-A implementation of the proposed model is presented. To achieve a reliable model, an approach to reproduce the phase noise generated by the MTJ STO has been proposed and successfully employed. The implemented model yields a time domain signal, which retains the characteristics of operating frequency, linewidth, oscillation amplitude and DC operating point, with respect to the magnetic field and applied DC current. The Verilog-A implementation is verified against the analytical model, providing equivalent device characteristics for the full range of biasing conditions. Furthermore, a system that includes an MTJ STO and CMOS RF circuits is simulated to validate the proposed model for system- and circuit-level designs. The simulation results demonstrate that the proposed model opens the possibility to explore STO technology in a wide range of applications.

\end{abstract}

\begin{IEEEkeywords}
spin torque oscillator, magnetic tunnel junction, macrospin, analytical model.
\end{IEEEkeywords}

\IEEEpeerreviewmaketitle

\section{Introduction}
\IEEEPARstart{S}{pintronics} is an emerging technology, utilizing both fundamental electronic charge and spin \cite{spintronics2001}. Spin is the intrinsic angular momentum of the electron. The spin-transfer-torque magnetoresistive random access memory (STT-MRAM), which is based on spintronic effects, has revolutionized the magnetic storage industry  \cite{STTMRAM1}-\cite{STTMRAM3}. In the past few years, extensive research on modeling this device has been carried out \cite{MTJ_model1}-\cite{MTJ_model3}. The developed models of STT-MRAM enable estimation of the performance of STT-MRAMs together with its CMOS circuits, further accelerating the development of STT-MRAM technology.

Meanwhile, the spin torque oscillator (STO) \cite{STObook}, which is another interesting spintronics-based device, has recently received a rapidly increased attention. The STO provides a widely tunable voltage oscillation at microwave frequencies, greatly extending the possible application range of spintronics. Possible applications of STOs include frequency detection \cite{nature2014}, \cite{tulapurkar2005spin}, magnetic field sensing \cite{nature2014}, \cite{braganca2010nanoscale}, microwave sources \cite{nature2014}, \cite{Bonetti2009}, \cite{Villard2010a} and microwave communications \cite{pufall2005frequency}, \cite{wireless2014}. Particularly, the magnetic tunnel junction (MTJ) STO, which provides comparatively large output power, show great potential to be used in different applications. This motivates the focus of this work on the MTJ STO. However, unlike for the case of STT-MRAM, very little progress has been achieved in modeling the STOs for circuit- and system-level design, impeding the development of STO-based applications. The only existing MTJ STO models \cite{MTJSTO_model1}, \cite{MTJSTO_model2} are limited by several factors. For instance, they offer a limited applicable range, inaccurate DC operating point, and utilize expressions that are not fully validated by experiments or theory. A new analytical MTJ STO model, which can overcome these issues, has been proposed in Part I \cite{Tingsu_2014TEDI}. 
To further allow being used by a circuit simulator, such as Cadence Spectre-RF, which can analyze the analog and RF performances of STO-based circuits and systems, the analytical MTJ STO model should be implemented in a hardware description language. 
Verilog-A is a hardware description language, which uses mathematical expressions to model the behaviors of arbitrary types of devices and components, while allowing \mbox{device-,} circuit- and system-level design and analyses.
Therefore, Verilog-A is suitable and will be used for modeling MTJ STOs.
In the existing MTJ STO models \cite{MTJSTO_model1}, \cite{MTJSTO_model2}, however, the information of full Verilog-A implementation is absent. 
In the STO Verilog-A model \cite{MTJSTO_model3}, which has been proposed by the same research group as \cite{MTJSTO_model1}, \cite{MTJSTO_model2} and has not been validated by MTJ STOs, it is not possible to change the bias magnetic field in the Verilog-A model since the calcuation of the effective magnetic field is not included. Instead, the effective magnetic field is calculated in advance in Matlab and manually imported to circuit simulation platforms by the user for every change of field bias condition.
Therefore, this model does not allow tuning the applied field in the circuit simulator, and it can not be considered to be fully implemented in Verilog-A, which brings difficulties in designing or optimizing the dedicated circuits for STOs.
Besides, to generate the frequency or phase fluctuation of the STO, the model in \cite{MTJSTO_model3} employs an approach that can cause signal discontinuity. As a result, this existing STO Verilog-A model is not ready to be used.

Here, in Part II, we present a full Verilog-A implemention of the anlytical MTJ STO model proposed in Part I \cite{Tingsu_2014TEDI}. This MTJ STO Verilog-A model uses a new approach to reproduce the phase fluctuation of the MTJ STO and avoid convergence issues, enabling a reliable MTJ STO model. Moreover, efficient simulations of both the stand-alone MTJ STO model and the MTJ STO model combined with CMOS circuits, are demonstrated. 

\section{Comprehensive and Compact MTJ STO Model in Verilog-A} 
As detailed in Part I \cite{Tingsu_2014TEDI}, the characteristics of an MTJ STO include the DC operating point, operating frequency $\omega_\text{g}$, output peak power $P(\omega)$, and linewidth $2 \Delta \omega$ (the full width at half-maximum). These characteristics vary greatly as the biasing condition changes. The biasing condition for the MTJ STO is typically composed of the amplitude ${H}_\text{ext}$ and the in-plane angle $\phi_\text{ext}$ of the external magnetic field, as well as the applied DC current  $I_\text{DC}$. 

The complete Verilog-A code of the comprehensive and compact MTJ STO model is available from \cite{model_link}.

\subsection{Computational efficiency of the comprehensive model}
To fully implement the MTJ STO model in Verilog-A, solving Eq.(4a, 4b) in \cite{Tingsu_2014TEDI} for the angle and magnitude of the effective field is realized solely by using Verilog-A. The effective field angle (see Eq.(4a) in \cite{Tingsu_2014TEDI}) is solved numerically using the fixed-point iteration method, based on which the effective field magnitude can be simply obtained using Eq.(4b) in \cite{Tingsu_2014TEDI}. The equation solver, as well as the the large amount of calculations involved in obtaining the nonlinear coefficients and characteristics of the MTJ STO, make the transient simulation time-consuming. For most of the analyses, none of the parameters in the effective field change for a single run, so that all the heavy calculations can be executed in the \emph{initial step} event in order to improve the efficiency of the simulation. The \emph{initial step} event is pre-defined in Verilog-A and called on the first point of a simulation. Therefore, in these cases, all the coefficients, parameters and characteristics are computed only once in the entire simulation. 

\subsection{Accurate phase generation}
In order to generate the time domain signal, which includes all the characteristics of the MTJ STOs, the output of the model should comprise of both RF and DC terms, implemented in Verilog-A as \\ 

\begin{tabular}{|l|}
\hline
$V$(MTJ\_STO, GND) <+ $V_\text{RF}+V_\text{DC};$ \\ \hline
\end{tabular} 
\\~
\\
where $V_\text{DC}$ is the DC voltage across the MTJ STO and is a function of the effective field angle, as detailed in \cite{Tingsu_2014TEDI}, and $V_\text{RF}$ can be derived based on Eq.(2) and Eq.(11) in \cite{Tingsu_2014TEDI}, written in Verilog-A as\\ 

\begin{tabular}{|l|}
\hline
$V_\text{RF}=R_\text{prec}\cdot I_\text{DC} \cdot \cos({\omega_0}\cdot \textdollar abstime+\varphi(t));$\\ \hline
\end{tabular} 
\\~
\\
where $R_\text{prec}$ is the amplitude of the resistance oscillation, whose value depends on the biasing condition. The $\textdollar abstime$ function returns the absolute simulation time. 
Since MTJ STOs have considerable phase noise amplitude, $\varphi(t)$ is used to represent the phase fluctuation (phase noise), reflecting the linewidth of the proposed model. For the timescales of circuit-level simulations (generally on the order of $\mu$s), white frequency noise extending over fluctuation frequencies wider than 1 MHz -- 100 MHz \cite{quinsat2010} will be the noise type dominating the linewidth. Hence, for this application, it is natural and adequate to approximate the frequency noise by white frequency noise alone. 
The amplitude noise of MTJ STOs is coupled to the phase noise due to the nonlinearity of the governing magneto-dynamic equation (Eq.(1) in \cite{Tingsu_2014TEDI}). Such a coupling greatly enhances the phase noise, making the impact of the amplitude noise less significant for applications \cite{quinsat2010}.  
For this reason, the amplitude noise is not explicitly addressed in this work, but its effect is manifested in the linewidth of the proposed MTJ STO model. Nevertheless, Gaussian amplitude noise could be added to the proposed model by further specifying its standard deviation. 

An accurate Verilog-A model of oscillators should memorize its phase along the simulation time. However, a solution to keep the phase information between adjacent simulation time steps has not been found. 
In one common Verilog-A phase noise generation method \cite{VCOmodel}, the step-specific perturbed frequency $f_{VCO}$+$\Delta f_{VCO}$ is integrated all the way from $t$ = 0 up to $t$, resulting in a discontinuous phase jump whenever $\Delta f_{VCO}$ is updated. The method used in the existing Verilog-A STO model \cite{MTJSTO_model3} instead implements a phase noise $\varphi_{STO}(t)$ that is discontinuously updated every couple of nanoseconds. In both methods, the comparatively large and discontinuous phase changes result in significant signal discontinuities for sinusoidal signals. This type of discontinuity is not a characteristic of MTJ STOs and, moreover, may cause convergence issues during simulations.

\begin{figure}[tb]
   \centering
  \begin{center}
    \includegraphics[trim = 5mm 34mm 7mm 30mm, clip, width=8.1cm]{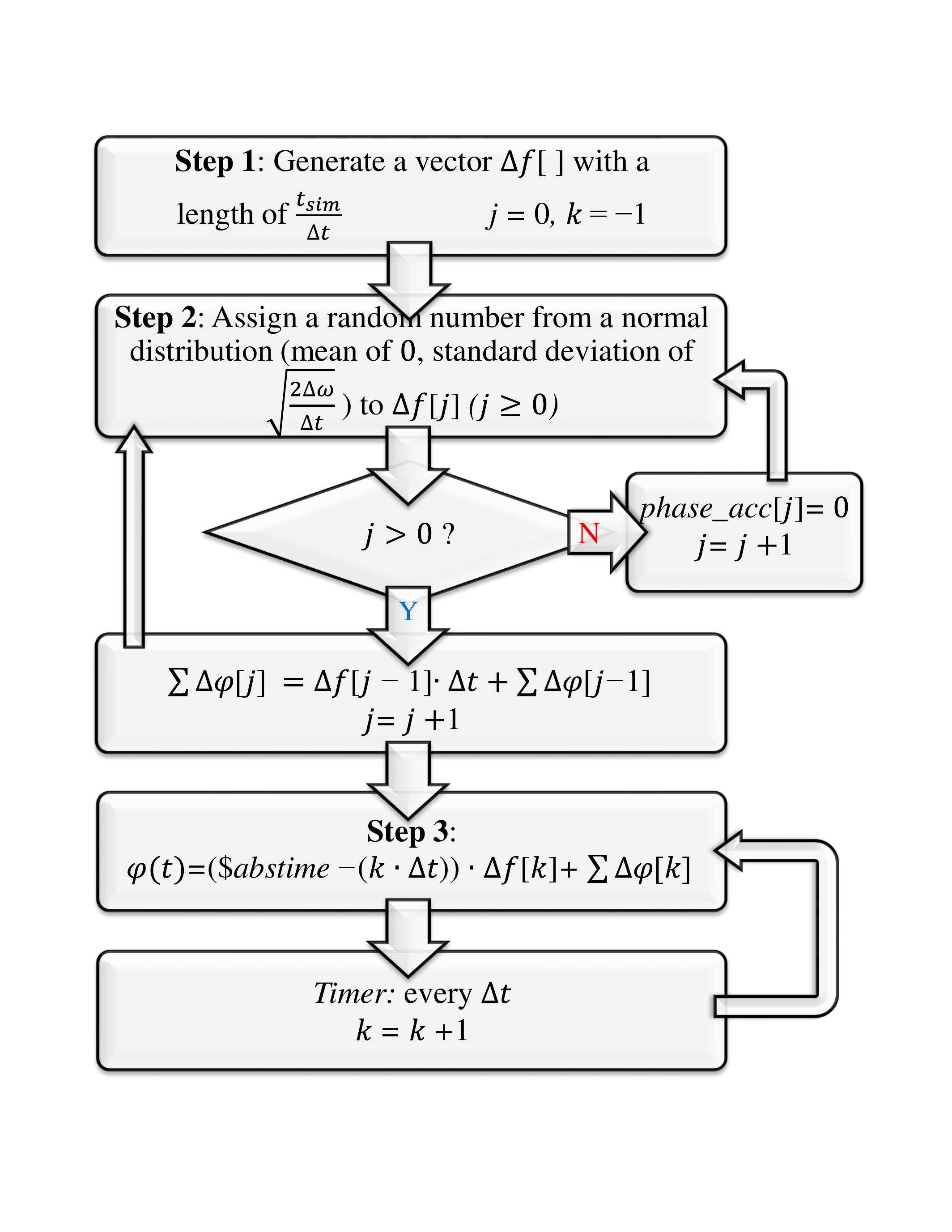}
    \centering
    \caption{Flow chart of $\varphi(t)$ implementation}
    \label{fig-label}
  \end{center}
\end{figure}
\begin{figure}[tb]
   \centering
  \begin{center}
    \includegraphics[trim = 37mm 119.5mm 67mm 78mm, clip, width=6.09 cm]{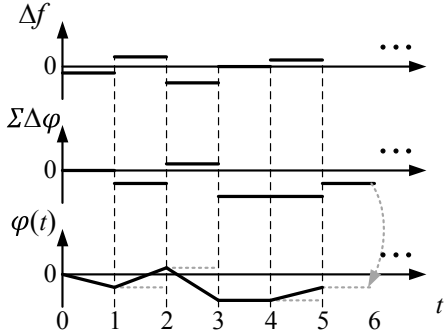}
    \centering
    \caption{Time domain illustration of implementing $\varphi(t)$ (the unit of the $\hat{x}$-axis is the virtual time step)}
    \label{fig-label}
  \end{center}
\end{figure}

The solution for achieving an accurate and reliable MTJ STO model is to instead implement a $\varphi(t)$, which bears a continuous, linear phase change in between the fixed points of randomized phase fluctuation. 
The flow chart of the divergence-free implementation of $\varphi(t)$ is illustrated in Fig. 1. The first step is to create a vector $\Delta f [~]$, which gives the random frequency fluctuation as a function of the linewidth (assumed that the frequency noise is white, i.e. that the linewidth has a Lorentzian lineshape), changing every $\Delta t$. $\Delta t$ is the virtual time step, which reflects how offen the frequency fluctuation happens and can be defined by the user. Moreover, the user-set $\Delta t$ is necessary in order to implement the Verilog-A model, because the circuit simulator provides an adaptive time step determined by the local truncation error (LTE) \cite{Kundert} so that the real simulation time step cannot be fully controlled by the user nor the programmer. Hence, the real simulation time step cannot be used to update the frequency fluctuations.
To obtain the correct linewidth, the dataset of $\Delta f [~]$ follows a normal distribution with a mean of 0 and a standard deviation of $\sqrt{\frac{2 \Delta \omega}{\Delta t}}$, resulting in a phase variance that is growing linearly with time in a rate consistent with the specific level of white frequency noise for the specific linewidth $2\Delta \omega$ \cite{silva2010}. The virtual time step $\Delta t$ is related to the upper cut-off frequency for white noise in the frequency noise spectrum, and should be set to a value smaller than one order of magnitude lower than the inverse cut-off frequency in order to produce white frequency noise all the way up to the cut-off. For cut-off frequencies of 100 MHz or 1 GHz, this means that the virtual time step $\Delta t$ should be set to lower than 1 ns or 100 ps respectively. Failing to set a short enough $\Delta t$ will result in a too narrow white band in the frequency noise spectrum, resulting in a decreased spectral linewidth. In this work, $\Delta t$ = 100 ps is employed.
The length of $\Delta f [~]$ is the ratio between the simulation time $t_{sim}$ and the virtual time step $\Delta t$. 

The second step in implementing $\varphi(t)$ is to create another vector $\Sigma\Delta \varphi[~]$, which is used to store the phase deviation accumulated from the reference time ($t$=0) to each virtual time step. The relationship between $\Delta f [~]$ and $\Sigma\Delta \varphi[~]$ is illustrated in Fig. 2. In the first period between $t$=0 and the first virtual time step ($t$=1), $\Sigma\Delta \varphi[j=0]$ is 0. 
During the period between $t$=$j$ and $t$=$j$+1 ($j$>0), the total accumulated phase deviation caused by the frequency fluctuation(s) in the past period(s) (from $\Delta f [0]$ to $\Delta f [j-1]$) is stored in $\Sigma\Delta \varphi[j]$ . 
A large $\Delta f [j]$ leads to a large phase deviation that will be fully presented at the next time step ($t$=$j$+1). The first two steps in implementing $\varphi(t)$ are necessarily conducted in the $initial$ $step$ event, so that the two vectors are generated only once during the simulation.

The third step, as shown in Fig. 1, is to generate the required $\varphi(t)$. 
$\varphi(t)$ at an arbitrary time, can be expressed as the sum of the accumulated phase deviation $\Sigma\Delta \varphi[k]$ up to the last virtual time step, and the phase deviation produced from the last virtual time step to the absolute time. 
The parameter $k$ is employed to count the past number of virtual time steps, and it is updated every $\Delta t$ by using the $timer$ function in Verilog-A. It should be noted that the initial value of $k$ is set to -1 since the $timer$ function is called at the beginning of each time period. The absolute time is fetched by calling the $\$ abstime$ function. The relationship between $\varphi(t)$, $\Delta f[k]$ and $\Sigma\Delta \varphi[k]$ is presented in Fig. 2. 
Specifically, the slope of $\varphi(t)$ is the instantaneous frequency deviation $\Delta f[k]$, so that the phase deviation generated from the last virtual time step to the absolute time is ($(\$ abstime - k \cdot \Delta t) \cdot \Delta f[k]$). By summing $\Sigma\Delta \varphi[k]$ and ($(\$ abstime - k \cdot \Delta t) \cdot \Delta f[k]$), the required $\varphi(t)$ is realized.

The proposed phase noise generation approach successfully overcomes the phase discontinuity issue identified in the existing Verilog-A models for oscillators \cite{MTJSTO_model3, VCOmodel}. However, the proposed noise generation approach makes the proposed MTJ STO model less suitable for simulations which involve momentary variations in the operating conditions, such as the modulation of current or field. 
Nevertheless, at the early stage of evaluating STOs in various applications and designing STO-based building blocks towards applications, these simulations involving momentary variations in the operating conditions are not yet critical.

\section{Simulation Results} 
Transient simulations of the aforementioned analytical model implemented in Verilog-A, are carried out using the Cadence SpectreRF circuit simulator. 

\subsection{Simulation results of the MTJ STO model}
To validate the proposed $\varphi(t)$ function for generating the frequency fluctuation of the MTJ STO, transient simulations of the stand-alone (unloaded) MTJ STO model using the parameters from different MTJ STOs \cite{ref1}-\cite{muduli2010nonlinear} are performed.  The time domain signals of the transient simulation using the parameters from \cite{ref1} and $I_\text{DC}$ = 1.5 mA, while sweeping $\phi_\text{ext}$, are depicted in Fig. 3 as an example. They are compared with the measured time domain signal from \cite{time}, since time domain measurements are not included in \cite{ref1}-\cite{muduli2010nonlinear}. 
The general nature of these simulated time domain signals agree with the measured one.
Especially, the phase fluctuation generated by the proposed $\varphi(t)$ function, has continuous changes and is very similar to that of the measured time domain signal \cite{time}. Aside from the discrepancies between the modeled (theoretical) and measured linewidth, this comparison demonstrates that the proposed MTJ STO model can reproduce the phase fluctuation generated by MTJ STOs, so as to achieve a reliable MTJ STO model. 

The time domain signals in Fig. 3. are further analyzed to validate that the model implemented in Verilog-A is equivalent to the analytical model given in \cite{Tingsu_2014TEDI}.
\begin{figure}[tb]
   \centering
  \begin{center}
    \includegraphics[trim = 47mm 106mm 24mm 40mm, clip, width=8.7cm]{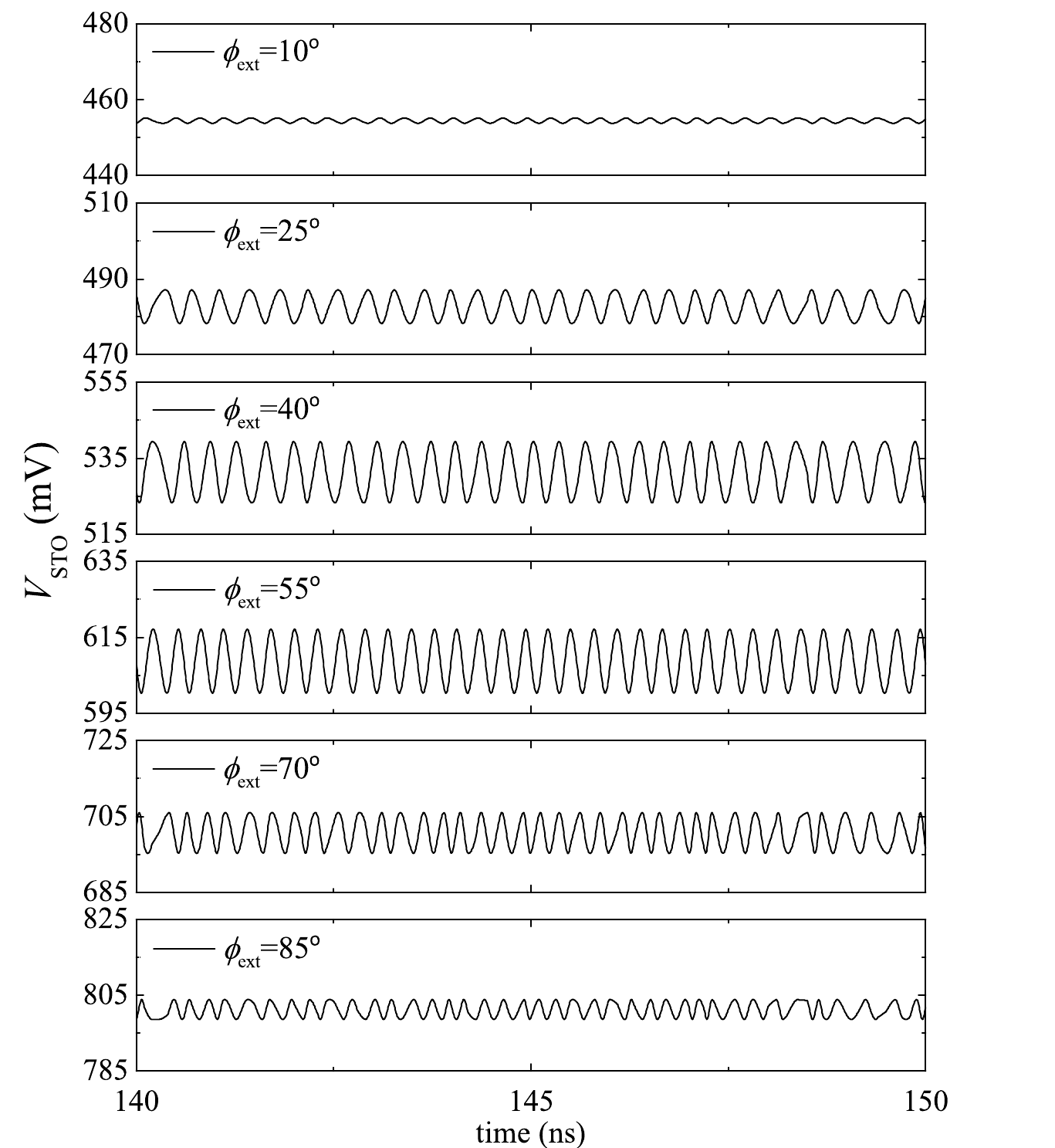}
    \centering
    \caption{Transient simulation results of the proposed MTJ STO model as a function of $\phi_{ext}$ ($I_\text{DC}$ = 1.5 mA)}
    \label{fig-label}
  \end{center}
\end{figure}
For different $\phi_\text{ext}$, the time domain signals shown in Fig. 3 present different characteristics including the operating frequency, phase noise (linewidth), amplitude, and DC biasing points. Time domain signals with large output power (voltage amplitude) appear at the in-plane external field angles between $40^o$ and $55^o$ (corresponding in-plane effective field angles between $47^o$ and $68^o$). Among these, the signal generated at $\phi_\text{ext}=55^o$ shows less random fluctuations in the phase than that at $\phi_\text{ext}=40^o$. The estimation of output power and linewidth based on Fig. 3 are in agreement with the analytical results given in \mbox{Fig. 3(a)} and Fig. 4(a) of \cite{Tingsu_2014TEDI} respectively.
As it can also be seen in Fig. 3, the phase noise at large in-plane external field angles degrades significantly, which is in accord with the theoretical linewidth in Fig. 4(a) of \cite{Tingsu_2014TEDI}.
In such cases, where the phase noise is considerable, the simulation does not suffer from any signal discontinuity or convergence issues, thanks to the proposed method used to generate the phase noise. 
Additionally, noticed from Fig. 3, the DC voltage across the MTJ STO increases as a function of $\phi_\text{ext}$, which is in agreement with Eq.(5) in \cite{Tingsu_2014TEDI}. 

To quantify the characteristics of the MTJ STO Verilog-A model as a function of $\phi_\text{ext}$, the time domain signals obtained from the transient simulations (1 $\mu$s) are converted (in Cadence) to the frequency domain using FFT so as to obtain power spectral densities (PSDs) of the signals, which are depicted in \mbox{Fig. 4(a)}. 
To perform the FFT, a Hamming window of the full waveform length 1 $\mu$s (16384 samples) is employed. This results in a resolution bandwidth of (1 $\mu$s)$^{-1}$ = \mbox{1 MHz}.
\begin{figure}[tb]
   \centering
  \begin{center}
    \includegraphics[trim = 28mm 79mm 25mm 34mm, clip, width=8.7cm]{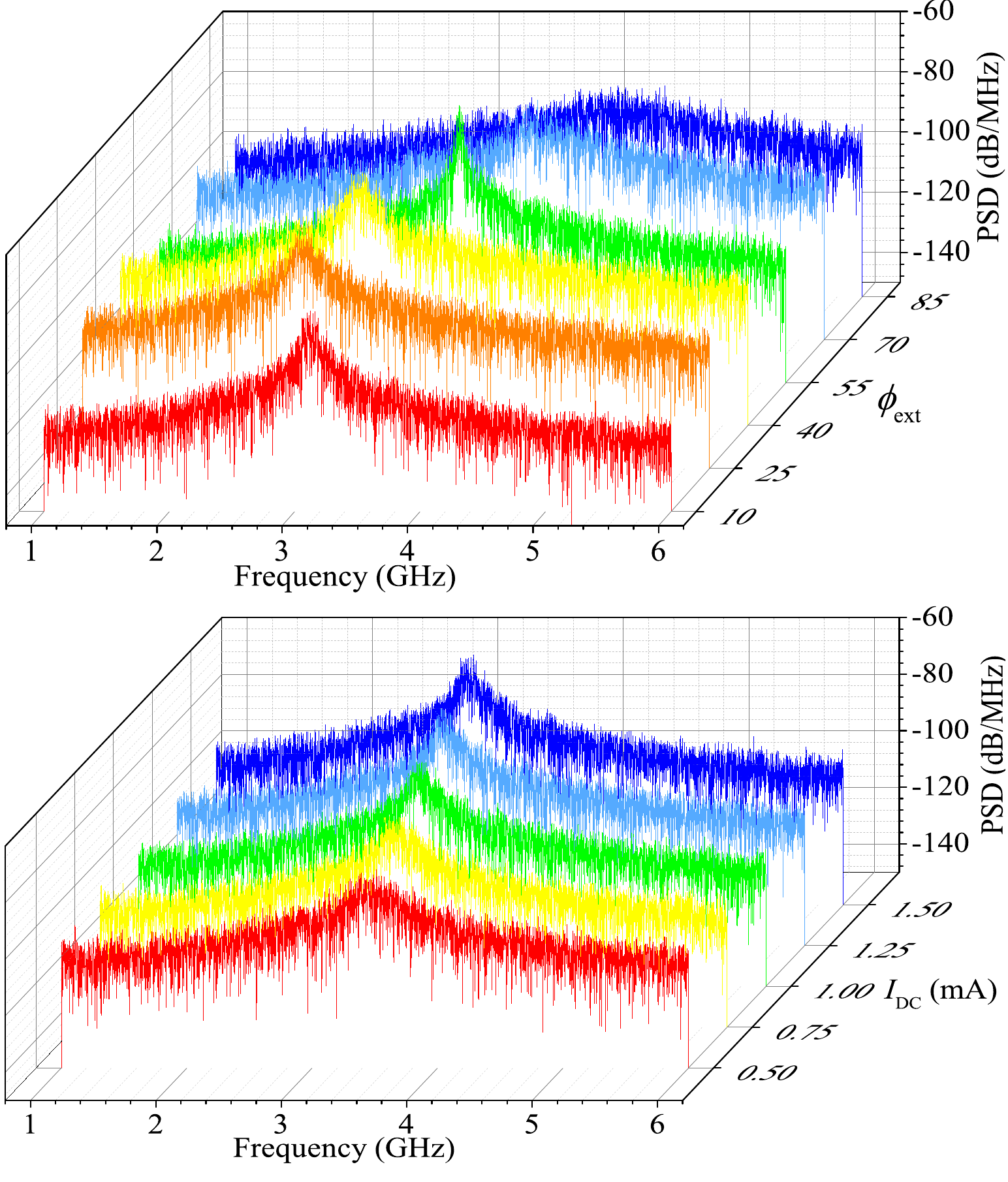}
    \centering
    \caption{PSDs of the time domain signals as a function of (a). $\phi_\text{ext}$ ($I_\text{DC}$ = 1.5 mA); (b). $I_\text{DC}$ ($\phi_\text{ext}= 45^o$)}
    \label{fig-label}
  \end{center}
\end{figure}

Fig. 4(a) shows the PSDs, which contain the information of the operating frequency and linewidth of the MTJ STO's signals as a function of $\phi_\text{ext}$. 
As $\phi_\text{ext}$ is increased, the operating frequency initially decreases, reaches a minimum and thereafter increases, agreeing with the measured data given in Fig. 2(a) of \cite{Tingsu_2014TEDI}.
For different $\phi_\text{ext}$, the operating frequency and linewidth of the proposed MTJ STO model implemented in Verilog-A are in accordance with the one obtained from the theoretical analysis, as given in Fig. 2(a) and Fig. 4(a) of \cite{Tingsu_2014TEDI}. Particularly, as it can be noticed from Fig. 4(a), the linewidth at $\phi_\text{ext}=55^o$, is much narrower than that at $\phi_\text{ext}=40^o$, indicating less random frequency fluctuations and confirming the estimation based on Fig. 3. Moreover, as it can be observed from Fig. 4(a), the signal with comparatively large output power can be found between $40^o$ and $55^o$, which is also in agreement with the analytical and measured results in Fig. 3(a) of \cite{Tingsu_2014TEDI}.

The dependence of $I_\text{DC}$ on the characteristics of the MTJ STO is also examined by performing transient simulations of the proposed Verilog-A model with different $I_\text{DC}$. The PSDs of the transient simulation results as a function of $I_\text{DC}$ at $\phi_\text{ext}=45^o$ is depicted in Fig. 4(b). The rise in $I_\text{DC}$ causes a decline in the operating frequency and an increase in the output power, which matches the theoretical results and experiments. 

Effort has been made in this work to achieve the compact MTJ STO model and improve the simulation speed. Therefore, the time required for transient simulations of the proposed model is of interest. To benchmark the simulation speed, a 1 $\mu$s simulation with a maximum time step of 5 ps is performed. By averaging the runtime of 10 simulations containing approximately 470900 transient steps, a simulation takes only 52.6 s. The simulations are performed on a server with 2$\times$AMD Opteron 6172, and occupy one core. Regarding the micromagnetics-based model, which is most efficiently simulated on graphic processing units (GPUs) or supercomputer clusters, the typical runtime is roughly 24 hours for a similar device. Compared with the micromagnetics-based model discussed in Part I \cite{Tingsu_2014TEDI}, the proposed model offers rapid simulations with a similar degree of accuracy.

\subsection{Hybrid simulation of the MTJ STO model with CMOS circuits}
Hybrid simulation of the MTJ STO model with CMOS circuits is of great importance since it provides validation of the proposed model at system- and circuit-level throughout its range of operation. 
Furthermore, it can fully cross-verify the model and the designed CMOS circuits. 
To perform the hybrid simulation, a system including the proposed MTJ STO model as well as CMOS circuits is considered. In this system, the MTJ STO model employs the parameters from \cite{ref3} since this MTJ STO has a resistance close to 50 $\Omega$, which eases analyses.
Firstly, proper biasing circuits for driving the MTJ STO are investigated and analyzed. 
For instance, the current mirror, which has been employed in \cite{MTJSTO_model1}, \cite{MTJSTO_model2} to provide the current biasing for the MTJ STO, is simulated with the proposed MTJ STO model and examined. 
The simulation results, however, suggest that using the current mirror to bias the MTJ STO is not suitable. The reason is that the resistance (biasing voltage) of the MTJ STO changes when $\phi_\text{ext}$ is varied, as illustrated in Fig. 3. 
Therefore, different resistances at different $\phi_\text{ext}$ make it impossible for the current mirror to accurately copy the current from a current source to the MTJ STO under all circumstances. 
Thereafter, a traditional RC bias-T, as shown in Fig. 5, is simulated with the proposed MTJ STO model. 
\begin{figure}[t]
   \centering
  \begin{center}
    \includegraphics[trim = 6mm 107mm 0mm 98mm, clip, width=9cm]{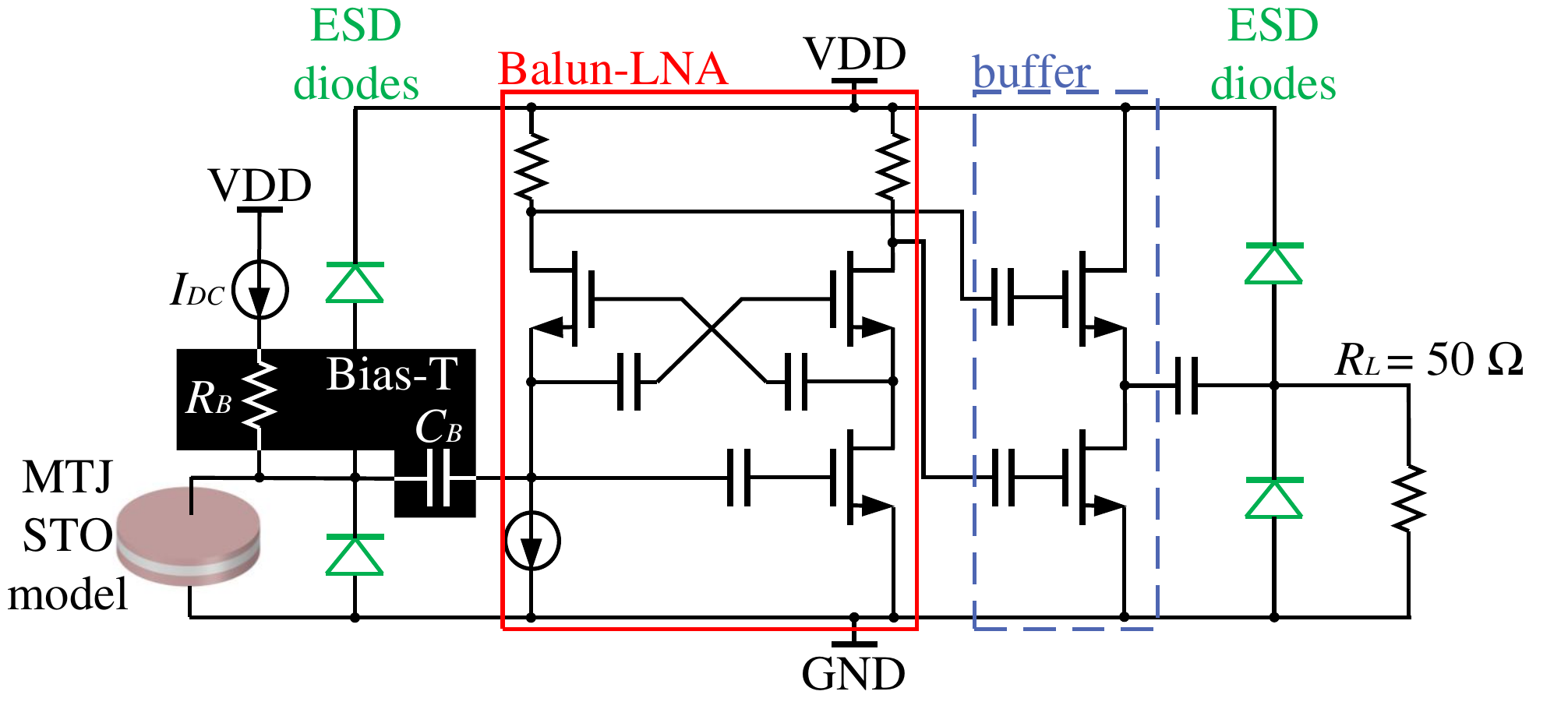}
    \centering
    \caption{The system comprising of a buffered Balun-LNA and the proposed MTJ STO model for hybrid simulations}
    \label{fig-label}
  \end{center}
\end{figure}
According to the simulation results, this RC bias-T can more successfully be utilized by MTJ STOs since the performance of the bias-T is not influenced by the variable resistance (biasing voltage) of MTJ STO. 
Consequently, the RC bias-T is used in this work to build the system. 
The selection of the biasing circuits for the MTJ STO demonstrates that the proposed MTJ STO model is very useful for the device and circuit community to identify the suitable circuit topologies and to design dedicated circuits for MTJ STOs, owing to the complete implementation of the proposed model in Verilog-A.

To complete a system, which provides low-noise amplification to MTJ STO signals and can be used in either applications or measurements, a wideband Balun-low noise amplifier (LNA) \cite{Tingsu2014} is employed. An output buffer and an AC coupling capacitor are added to present a system (Fig. 5) that is able to drive 50 $\Omega$ load.
The buffered wideband Balun-LNA is fully-ESD protected, and it is implemented in CMOS 65 nm process with a 1.2 V power supply. 

Before performing the transient simulation of the system, the unloaded MTJ STO is simulated at \mbox{$I_\text{DC}= 3$ mA} and $\phi_\text{ext}=40^o$, where a comparatively high voltage generated by the MTJ STO can be obtained. 
This voltage will be used as the reference voltage to compare with the voltage signals obtained from the MTJ STO together with the Balun-LNA. 
The transient simulation of the system is then conducted at the same biasing condition.
The obtained time domain signals, including the voltage generated by the unloaded MTJ STO, the voltage that can be delivered from the MTJ STO to the Balun-LNA, and the amplified voltage at the differential output of the Balun-LNA, are plotted in Fig. 6. 
\begin{figure}[t]
   \centering
  \begin{center}
    \includegraphics[trim =46mm 106mm 27mm 113.5mm, clip, width=8.7cm]{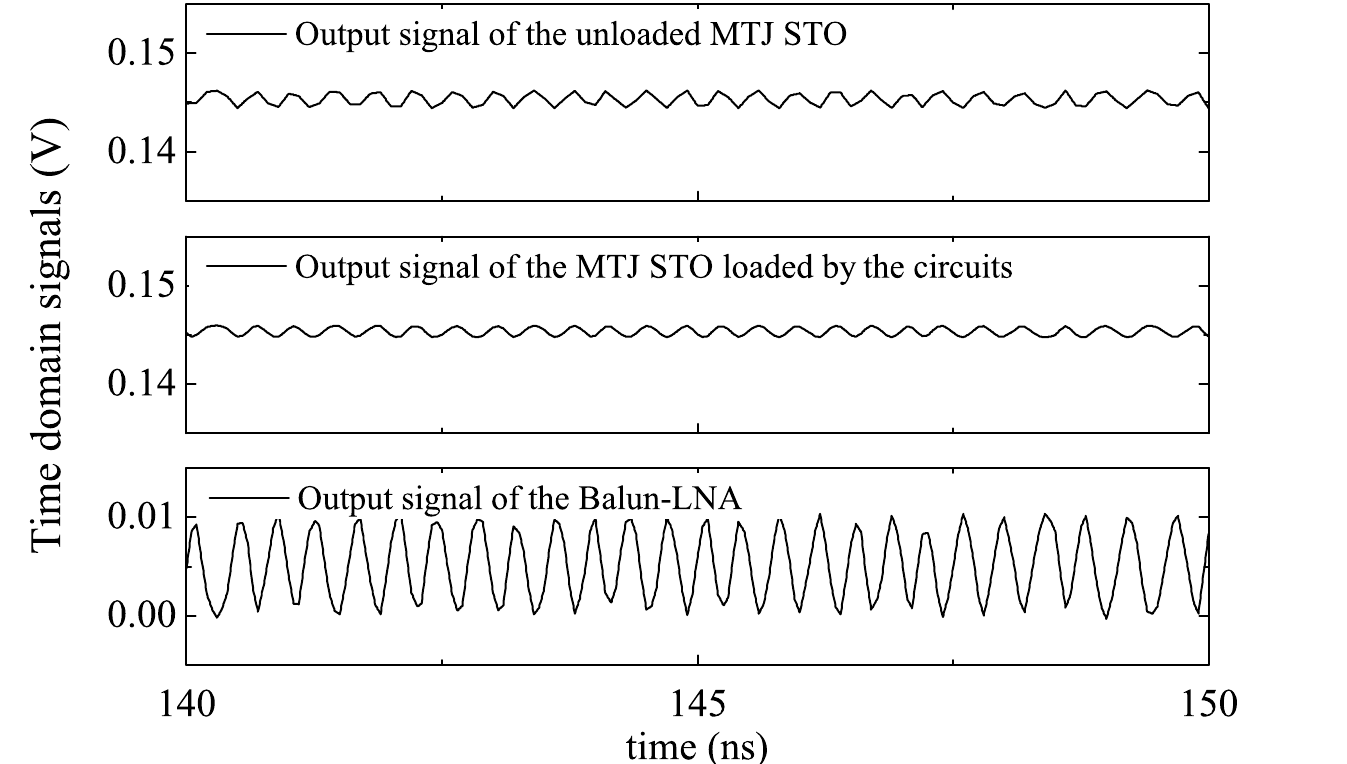}
    \centering
    \caption{Transient simulation results of the CMOS Balun-LNA with the MTJ STO model ($I_\text{DC}$ = 3 mA, $\phi_\text{ext}= 40^o$)}
    \label{fig-label}
  \end{center}
\end{figure}
The output signals of the MTJ STO before and after being connected by the Balun-LNA show that the DC voltage is sustained due to employed bias-T, and the DC resistance ($R_\text{DC}$) is close to 50 $\Omega$. In addition, approximately 2/3 of the AC voltage generated by the MTJ STO is delivered to the Balun-LNA, which is due to the fact that an AC voltage divider is formed between $R_\text{DC}$ and the input impedance of the Balun-LNA ($Z_\text{in}$). 
To quantify the power delivery and amplification so as to evaluate the hybrid simulation, PSDs of these signals are plotted in Fig. 7. 
\begin{figure}[t]
   \centering
  \begin{center}
    \includegraphics[trim =24mm 162mm 30mm 36mm, clip, width=8.7cm]{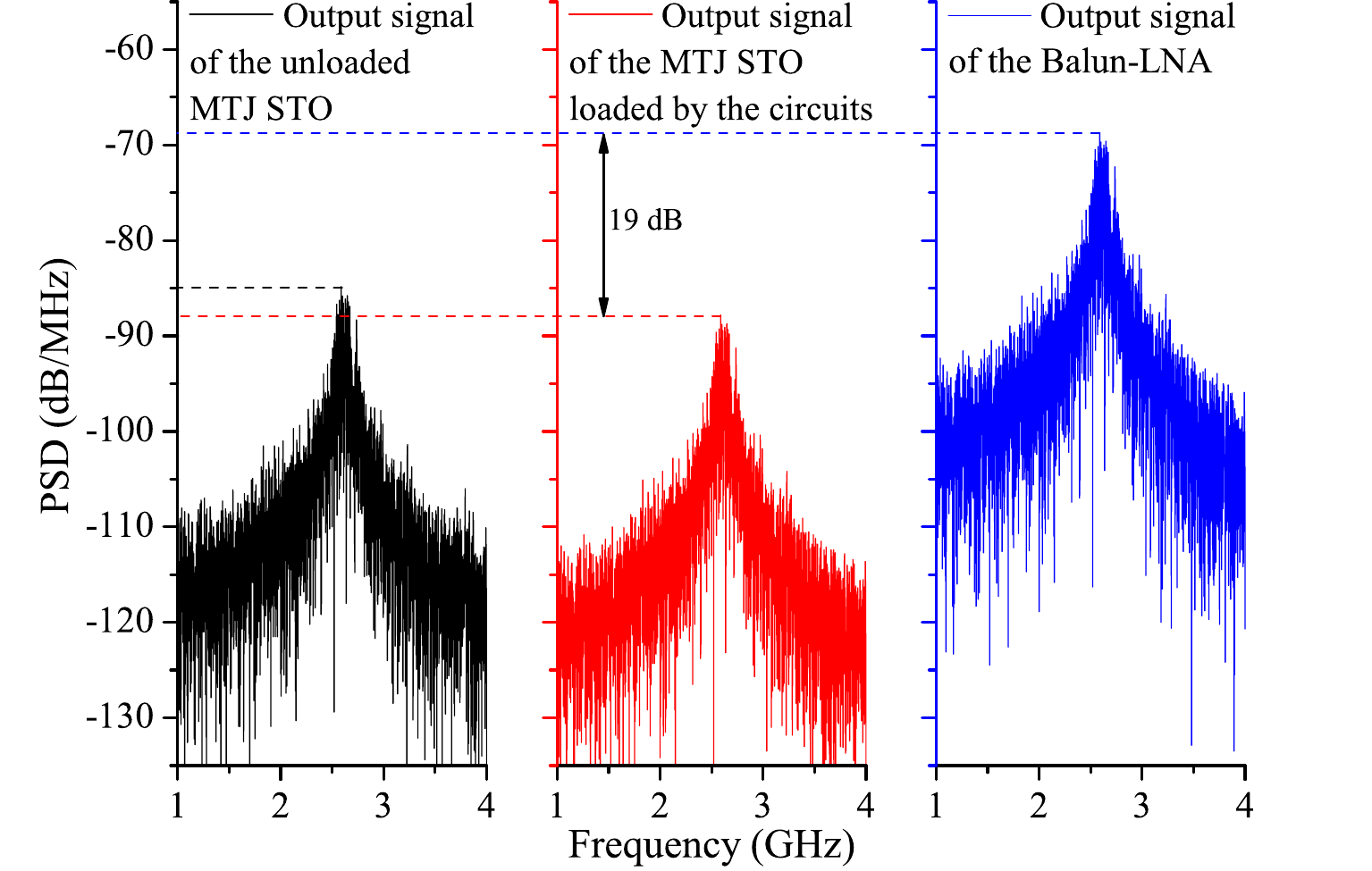}
    \centering
    \caption{PSDs of the signals obtained during the hybrid simulation of the MTJ STO model with the CMOS Balun-LNA}
    \label{fig-label}
  \end{center}
\end{figure}
Figure 7 shows that the power delivered from the MTJ STO to the Balun-LNA is about -3 dB less than the power generated by the unloaded MTJ STO. 
Based on this \mbox{-3 dB} loss ($Z_\text{in}$/($Z_\text{in}+R_\text{DC}$) = 10$^{\frac{-3}{20}}$), it can be calculated that the return loss S11=($Z_\text{in}-R_\text{DC}$)/($Z_\text{in}+R_\text{DC}$) at the operating frequency of the MTJ STO (under the applied biasing condition) is \mbox{-7.5 dB}. Since $R_\text{DC}$ is close to 50 $\Omega$, the calculated S11 should be similar to the S11 of the Balun-LNA that is characterized with a 50 $\Omega$ termination. The S11 of the Balun-LNA reported in \cite{Tingsu2014} is about \mbox{-8 dB}, which verifies the S11 calculated based on the hybrid simulation.   
As it can be also approximated from Fig. 7, the difference between the simulated PSD signals at the input and output of the Balun-LNA is approximately \mbox{19 dB}, which corresponds to the gain of the Balun-LNA reported in \cite{Tingsu2014}.
In summary, the behavior of the proposed MTJ STO model has been validated at circuit- and system- level.
The evaluated hybrid simulation demonstrates that the performance of an MTJ STO-based system can be easily, reliably and accurately predicted by the circuit simulator using the proposed MTJ STO Verilog-A model. 

\section{Conclusion} 
The analytical model of the MTJ STO proposed in \mbox{Part I} of this paper has been fully implemented in Verilog-A, enabling its direct use in STO-based systems. During the implementation, an approach to replicate the phase noise, hence the generated signal of the MTJ STOs, has been developed. This approach makes a reliable MTJ STO model possible, and allows different performance analyses so as to extensively explore possible applications. The simulation results of the stand-alone MTJ STO model and the MTJ STO-based system show that the implemented model gives identical characteristics as those obtained from the proposed analytical model. 
Additionally, the results demonstrate that the proposed MTJ STO model is useful for estimating as well as improving overall performance of the MTJ STO-based circuits and systems. Consequently, the proposed MTJ STO has the potential to accelerate the development of MTJ STO technology towards its future applications.

\ifCLASSOPTIONcaptionsoff
  \newpage
\fi

\bibliographystyle{plain}

\begin{IEEEbiography}[{\includegraphics[trim = 25mm 117mm 72mm 40mm width=1in,height=1.25in,clip,keepaspectratio]{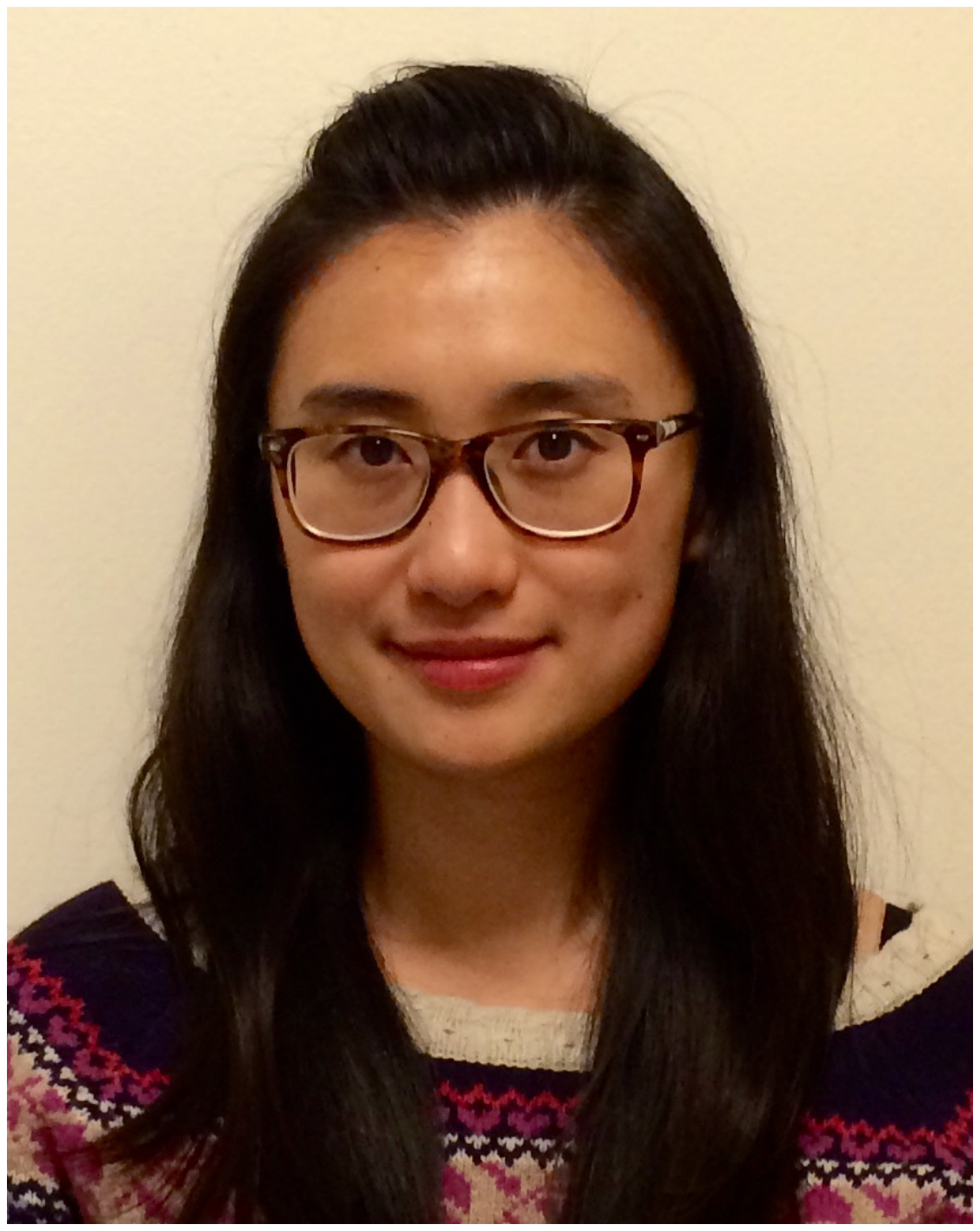}}]
{Tingsu Chen}
(S'11) received the B.Sc. degree in communication engineering from the Nanjing University of Information Science and Technology, China, and the M.Sc. degree in system-on-chip design from the KTH Royal Institute of Technology, Sweden, in 2009 and 2011, respectively. She is currently working toward the Ph.D. degree at KTH with the research area of high frequency circuit design for spin torque oscillator technology.
\end{IEEEbiography}
\begin{IEEEbiography}[{\includegraphics[trim = 25mm 117mm 72mm 40mm width=1in,height=1.25in,clip,keepaspectratio]{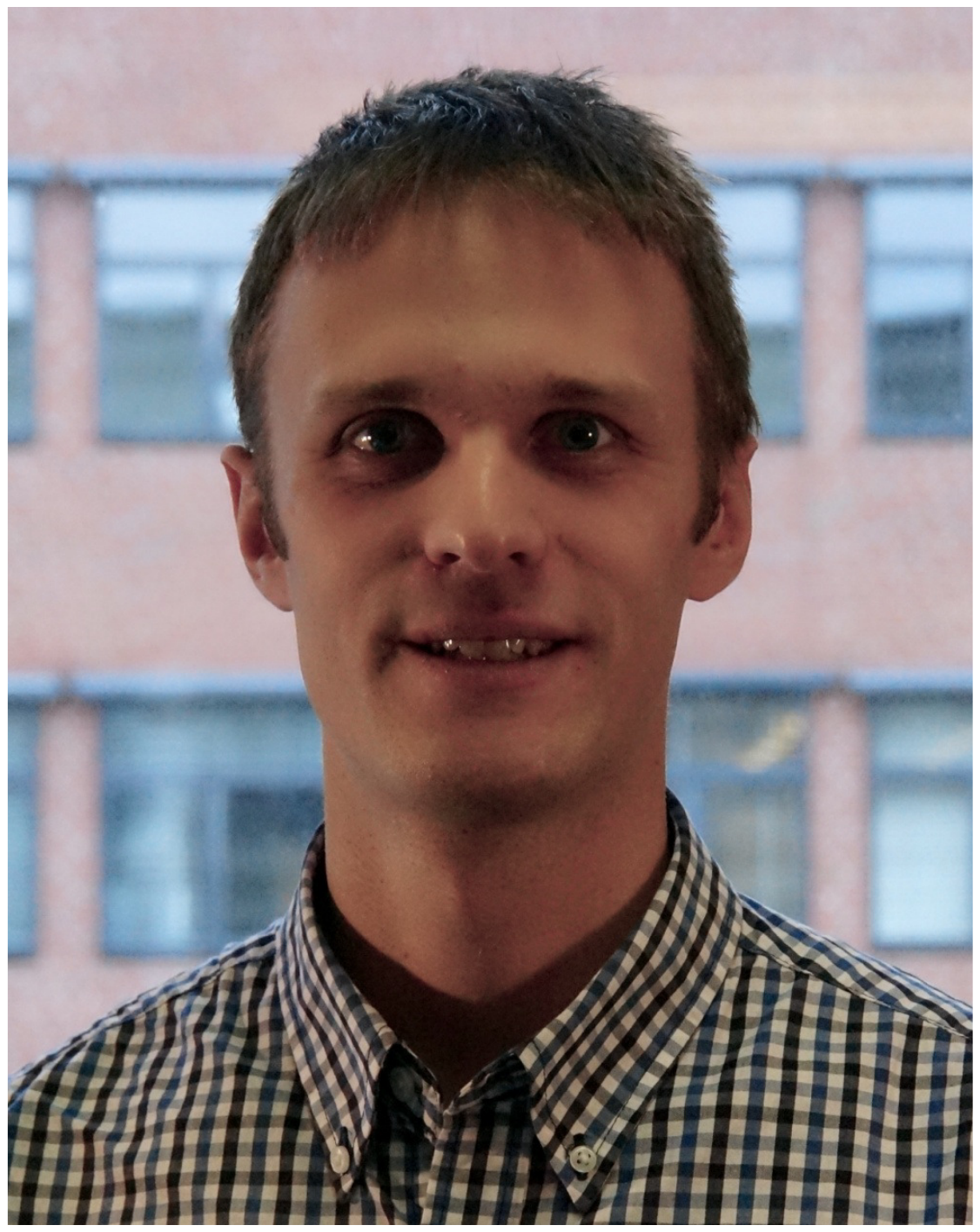}}]
{Anders Eklund}
(S'13) received the M.Sc. degree in engineering physics from KTH Royal Institute of Technology, Sweden, in 2011. He is currently working towards a Ph.D. degree in physics at KTH, experimentally investigating the frequency stability of spin torque oscillators by means of electrical characterization and synchrotron x-ray measurements.
\end{IEEEbiography}
\begin{IEEEbiography}[{\includegraphics[trim =  25mm 117mm 72mm 40mm width=1in,height=1.25in,clip,keepaspectratio]{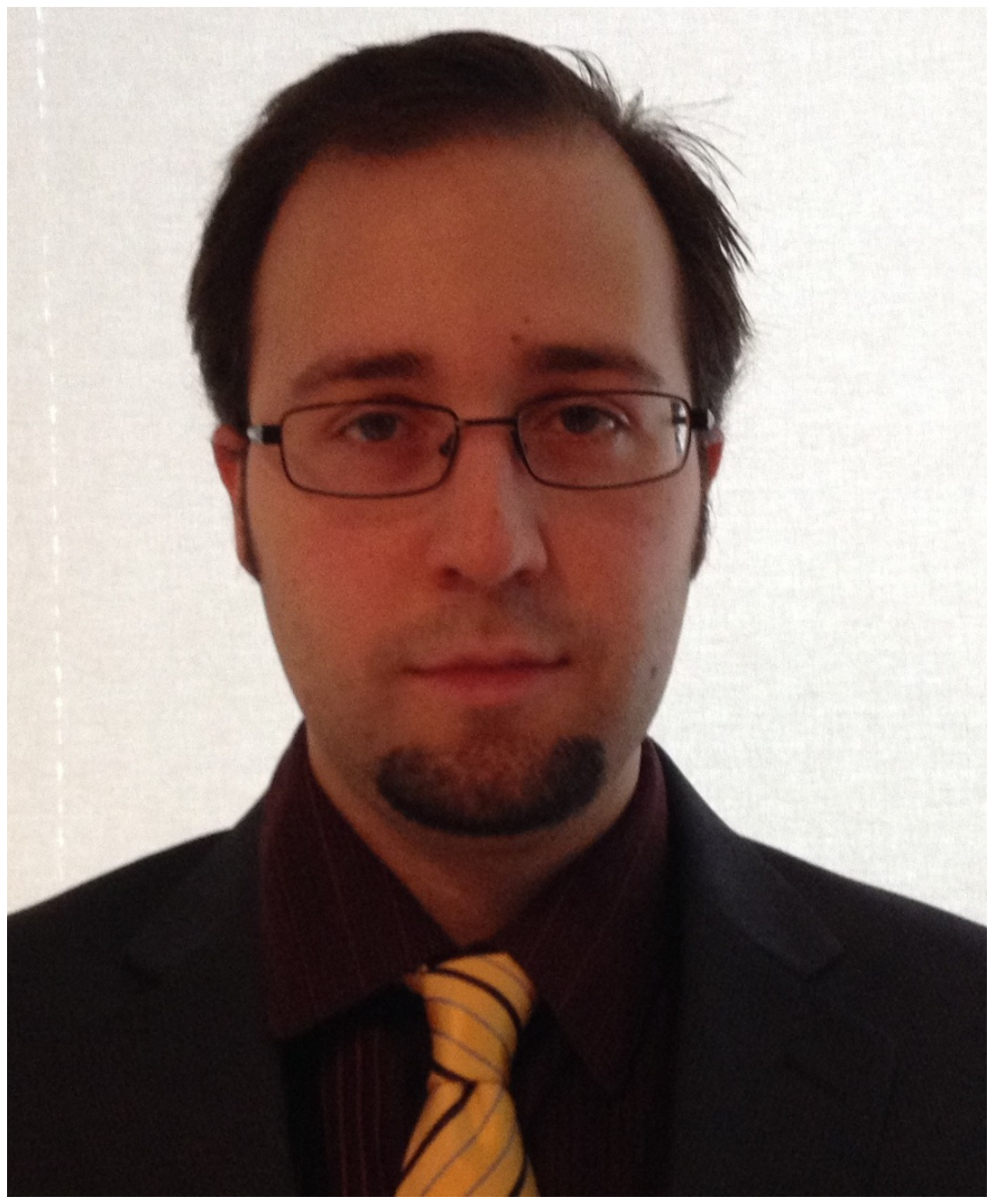}}]
{Ezio Iacocca}
(S'08) received the B.Sc. degree in electronic engineering from the Sim\'{o}n Bol\'{i}var University, Caracas, Venezuela ('08), the M.Sc. in nanotechnology from the Royal Institute of Technology, Stockholm, Sweden ('10), and the Ph.D. in physics from the University of Gothenburg, Gothenburg, Sweden ('14). His research focuses on the magnetodynamical modes of spin transfer torque driven nano oscillators and their applications in communication and storage technology.
\end{IEEEbiography}
\begin{IEEEbiography}[{\includegraphics[trim =  25mm 117mm 72mm 40mm width=1in,height=1.25in,clip,keepaspectratio]{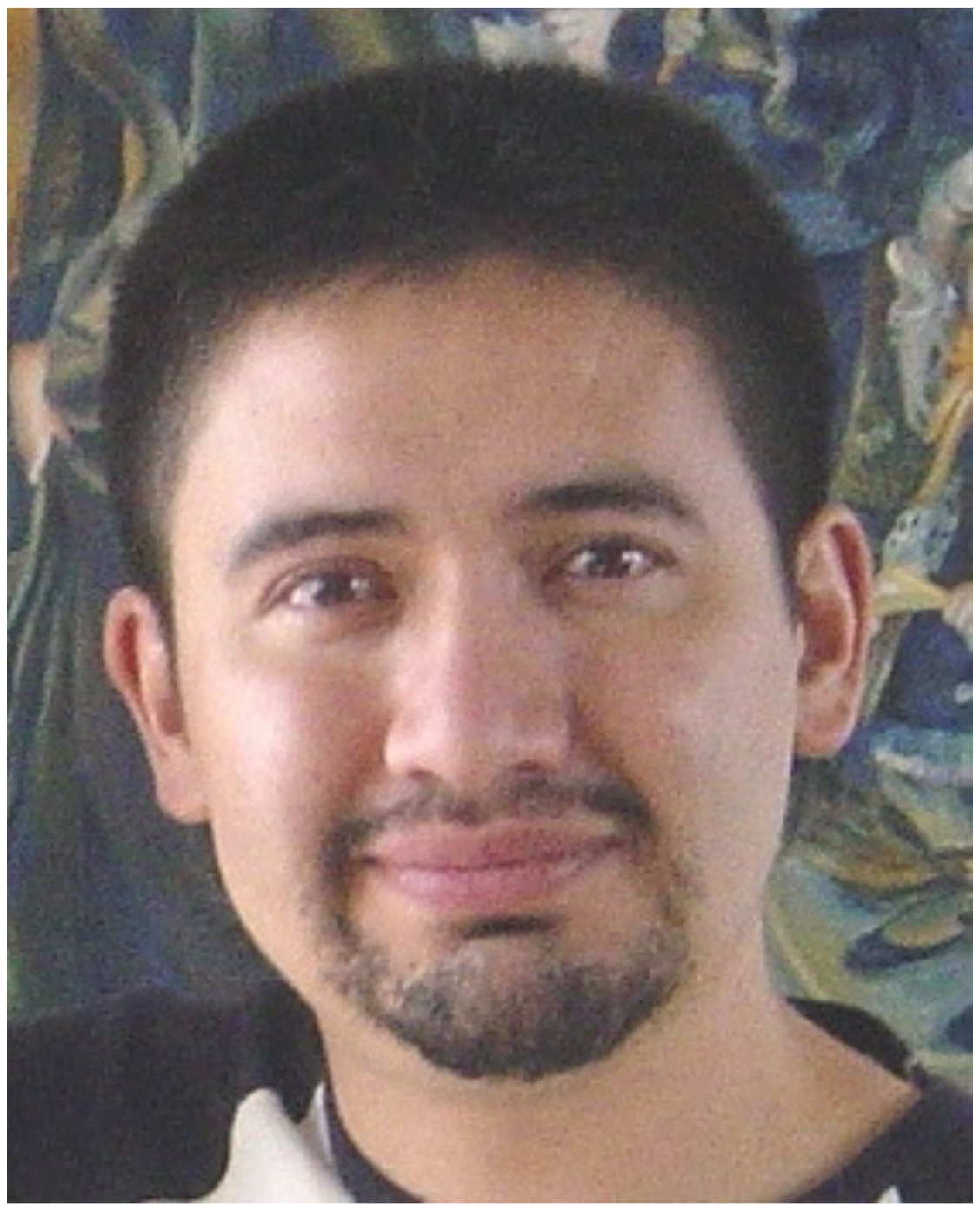}}]
{Saul Rodriguez}
(M'06) received the B.Sc. degree in electrical engineering from the Army Polytechnic School (ESPE), Quito, Ecuador, and the M.Sc. degree in system-on-chip design and the Ph.D. degree in electronic and computer systems from the KTH Royal Institute of Technology, Stockholm, Sweden. in 2001, 2005, and 2009, respectively. His current research interests include RF CMOS circuit design for wideband frond-ends, ultralow-power circuits for medical applications and graphene-based RF, and AMS circuits.
\end{IEEEbiography}
\begin{IEEEbiography}[{\includegraphics[trim =  25mm 117mm 72mm 40mm width=1in,height=1.25in,clip,keepaspectratio]{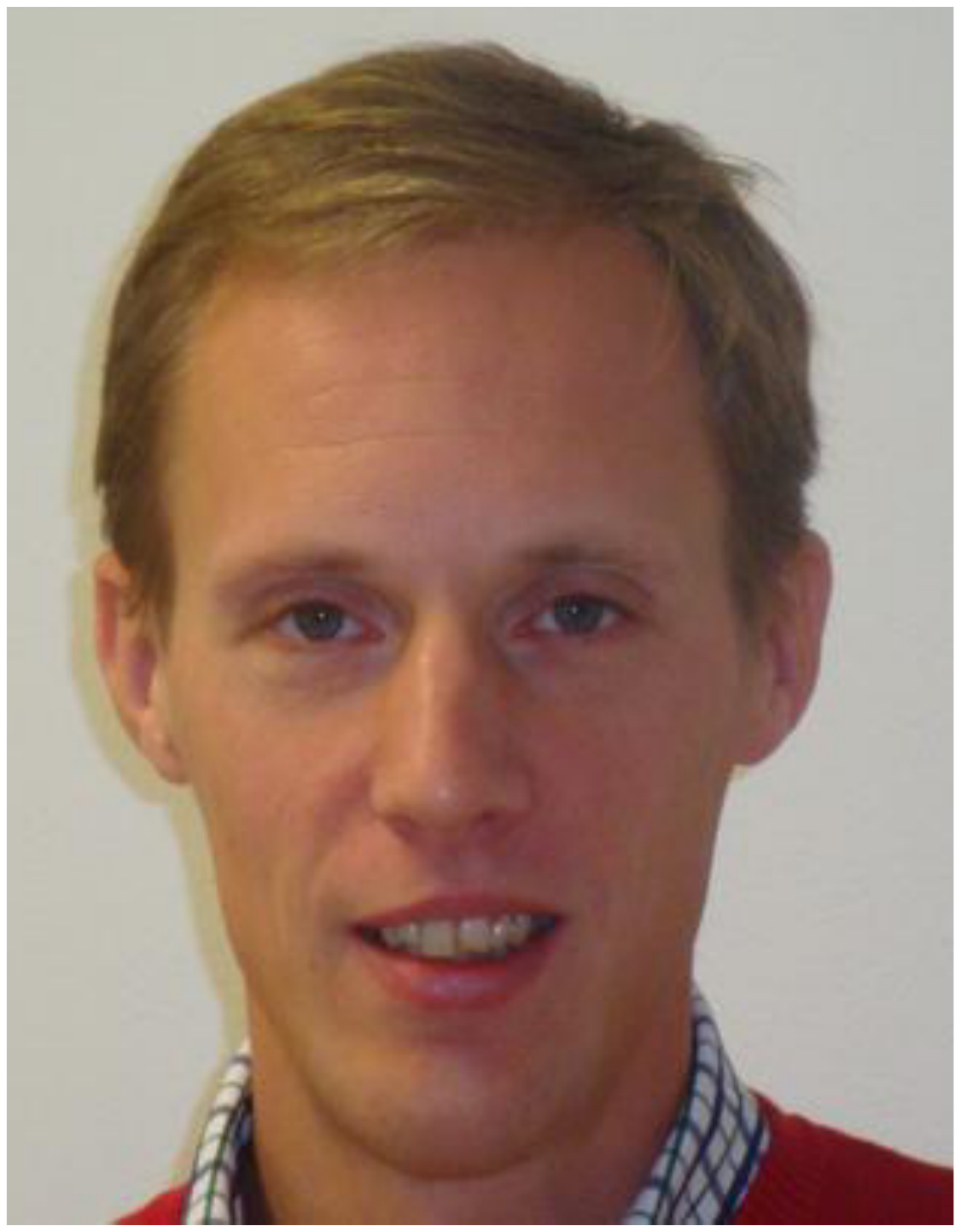}}]
{B. Gunnar Malm}
(M'98 - SM'10) was born in Stockholm, Sweden, in 1972. He received the M.S. from Uppsala University, Sweden, in 1997, the PhD in solid-state electronics 2002, from Royal Institute of Technology (KTH), Stockholm. He is an Associate Professor at the School of ICT, KTH since 2011. His recent work includes silicon photonics, silicon-carbide technology for extreme environments and spintronics. He also serves on the KTH Sustainability Council.
\end{IEEEbiography}
\begin{IEEEbiography}[{\includegraphics[trim =  25mm 117mm 72mm 40mm width=1in,height=1.25in,clip,keepaspectratio]{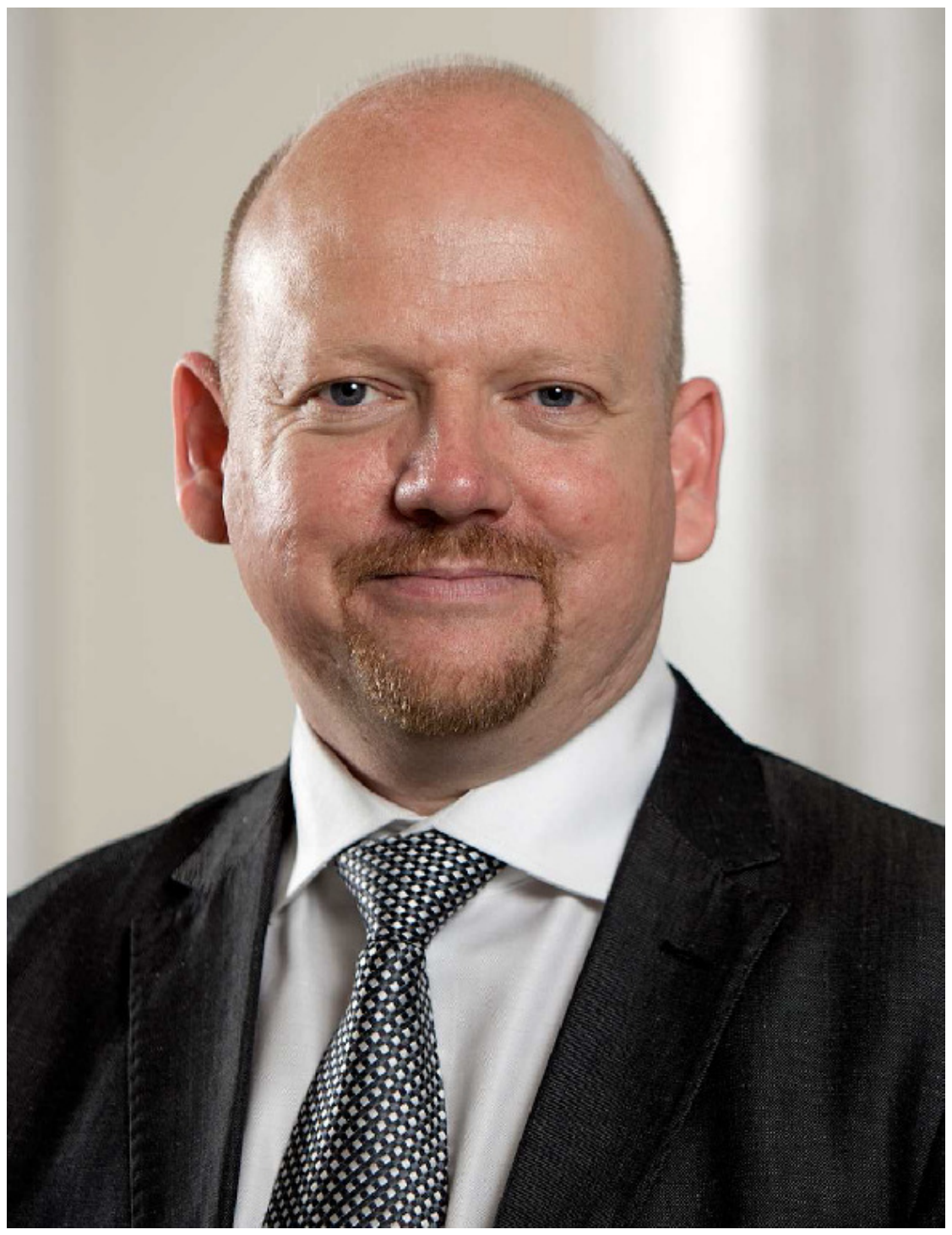}}]
{Johan $\AA$kerman}
(M'06) has an Ing. Phys. Dipl. degree ('94) from EPFL, Switzerland, a M.Sc. in physics ('96) from LTH, Sweden, and a Ph.D. in materials physics ('00) from KTH Royal Institute of Technology, Stockholm. In 2008 he was appointed Full Professor at University of Gothenburg and is a Guest Professor at KTH Royal Institute of Technology. He is also the founder of NanOsc AB and NanOsc Instruments AB.
\end{IEEEbiography}
\begin{IEEEbiography}[{\includegraphics[trim =  25mm 117mm 72mm 40mm width=1in,height=1.25in,clip,keepaspectratio]{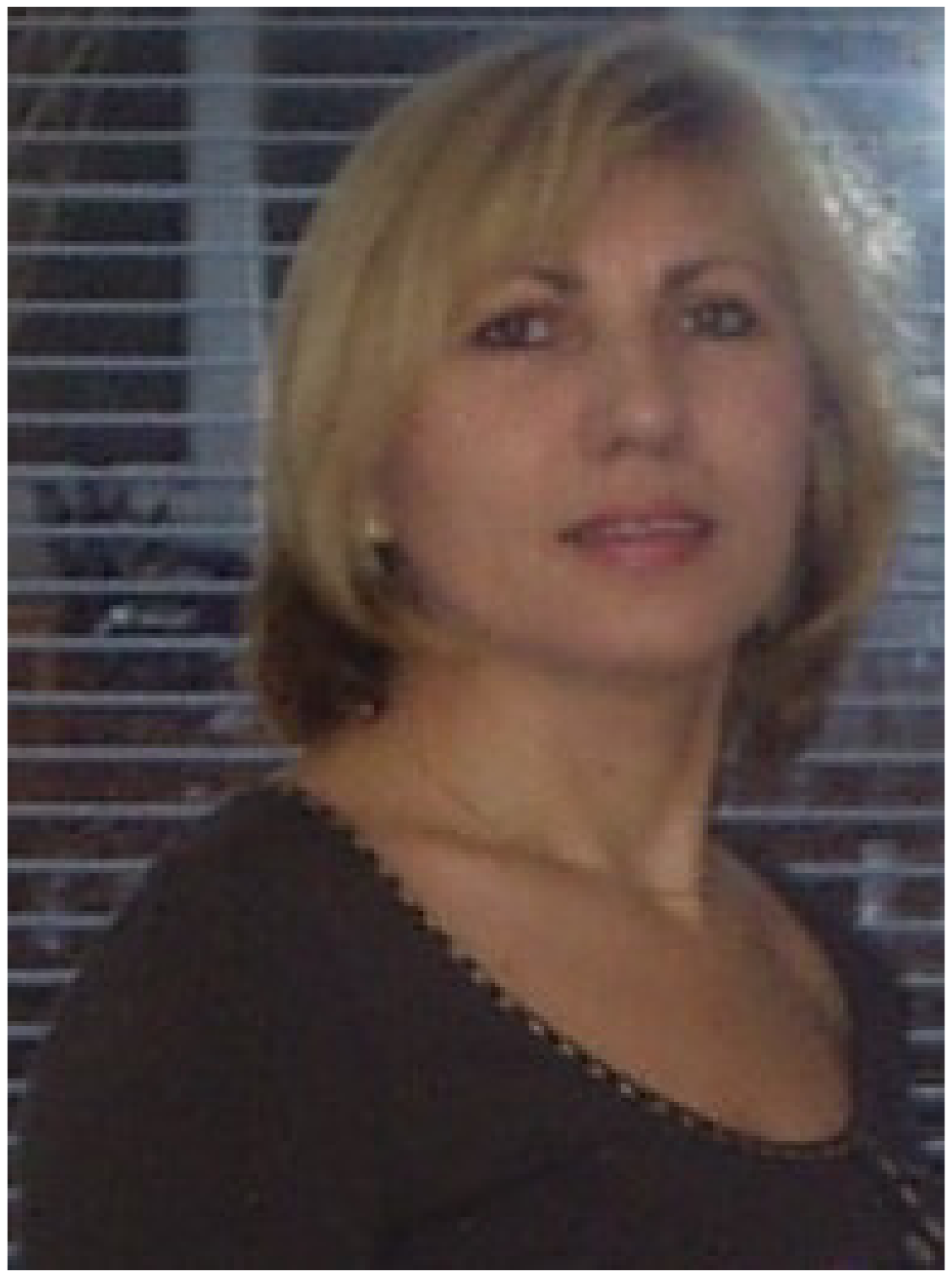}}]
{Ana Rusu}
(M'92) received the M.Sc. degree in electronics and telecommunications and Ph.D. degree in electronics in 1983 and 1998, respectively. She has been with KTH Royal Institute of Technology, Stockholm, Sweden, since 2001, where she is Professor in electronic circuits for integrated systems. Her research interests include low/ultralow power high performance CMOS circuits and systems, STO-based systems, RF graphene and high temperature SiC circuits.
\end{IEEEbiography}
\vfill 

\end{document}